\documentclass[twocolumn,amsmath,floatfix,prb,aps]{revtex4-1}
\usepackage{color}
\usepackage{amsmath}
\usepackage{pifont}   
\usepackage{graphicx} 
\usepackage{dcolumn}  
\usepackage{bm}       
\usepackage{amsfonts} 
\usepackage{amssymb}  
\usepackage{multirow} 

\usepackage[english]{babel}
\usepackage[autostyle, english = american]{csquotes}
\MakeOuterQuote{"}

\usepackage{placeins}

\usepackage{braket} 

    \renewcommand{\v}[1]{\bm{\mathrm{#1}}}
    \newcommand{\m}[1]{\bm{\mathsf{#1}}}

\begin{document}

\title{Numerical analysis of XMCD sum rules at the $L$-edge: when do they fail?}
\author{S. Shallcross$^1$}
\author{J.~K. Dewhurst$^2$}
\author{P. Elliott$^1$}
\author{S. Eisebitt$^{1,3}$}
\author{C. von Korff Schmising$^1$}
\author{S. Sharma$^1$}
\email{sharma@mbi-berlin.de}
\affiliation{1 Max-Born-Institute for Non-linear Optics and Short Pulse Spectroscopy, Max-Born Strasse 2A, 12489 Berlin, Germany}
\affiliation{2 Max-Planck-Institut fur Mikrostrukturphysik Weinberg 2, D-06120 Halle, Germany}
\affiliation{3 Institute for Optics and Atomic Physics, Technische Universit\"at Berlin,  10623 Berlin, Germany}

\date{\today}

\begin{abstract}
In the highly non-equilibrium conditions of laser induced spin dynamics magnetic moments can only be obtained from the spectral information, most commonly from the spectroscopy of semi-core states using the so-called x-ray magnetic circular dichroism (XMCD) sum rules. The validity of the these sum rules in tracking femtosecond spin dynamics remains, however, an open question.
Employing the time dependent extension of density functional theory (TD-DFT) we compare spectroscopically obtained moments with those directly calculated from the TD-DFT densities.
We find that for experimentally typical pump pulses these two very distinct routes to the spin moment are, for Co and Ni, in excellent agreement, validating the experimental approach. 
However, for short and intense pulses or high fluence pulses of long duration the XMCD sum rules fail, with errors exceeding 50\%. This failure persists only during the pulse and occurs when the pump pulse excites charge out of the $d$-band and into $sp$-character bands, invalidating the semi-core to $d$-state transitions assumed by the XMCD sum rules.
\end{abstract}

\maketitle


{\it Introduction}: The ultrafast control over magnetism by light\cite{bigot1996} offers a paradigm shift away from a primacy of charge excitations towards one of controlled spin excitations, with the promise of profoundly more energy efficient memory storage technologies. Central to progress in this field is the ability to reliably measure the evolution of spin moments on femtosecond timescales\cite{bovensiepen2009}, which for such strongly out of equilibrium systems can be obtained via spectral information. One of the most well established experimental approach by which this is achieved is through magnetic circular dichroism (MCD) and x-ray absorption spectroscopy (XAS), which can be combined via certain sum rules to obtain element specific spin and orbital angular momenta on ultrafast time scales\cite{boeglin2010,stamm2010,bergeard2014, Hennecke2019a}. Whether magnetic moments derived from such response functions (XMCD and XAS) are in agreement with the fundamental magnetic moments in solids remains, however, an ongoing source of controversy\cite{resta2020,altarelli2020}.

The XMCD sum rules are derived within an atomic picture in which the probing light pulse causes transitions from deep semi-core states to well defined atomic-orbitals\cite{thole1992,carra1993,chen1995,altarelli1993,wu1994,resta2020,altarelli2020,kunes2000}. In the ground state the moment of many magnetic materials is carried predominately by quasi-particles of $d$ (transition metals) or $f$ (lanthanides) character, and so this assumption can be expected to hold good, given that these orbitals are highly localized in nature. However, under laser pump conditions the electronic structure can dramatically alter, with ground state tightly bound electrons delocalizing into high angular momentum states. Such leakage from the $d$- or $f$-band into higher $l$ states would invalidate the key assumption underpinning the XMCD sum rules, throwing into question the use of XMCD in probing magnetism at femtosecond time scales\cite{carva2009,resta2020,altarelli2020}.

Theoretically, to answer this question (are the XMCD sum-rules valid for solids highly out of equilibrium?) one needs to time propagate the Hamiltonian under pump pulse conditions and directly compare the two routes for determination of the magnetic moment: (a) calculation of the transient L-edge XMCD and XAS spectra, and then using the sum-rules, exactly as in experiments, to  obtain the spin and the orbital moments and (b) the spin and orbital angular momentum obtained from the wave function by calculating the expectation value of the corresponding operators. 
Good agreement between the two would indicate that even for the highly non-equilibrium state of an extended solid (in which bands lose their atomic character), the MCD sum rules are a reliable measure of the transient magnetic moment. This approach has not been attempted to date, largely due to the formidable numerical difficulty entailed: it requires extensive calculations using highly precise all electron methods, with the latter mandatory as one must treat on an equal footing both states as low as 870~eV (2$p$-states) as well as states around the Fermi energy ($d$-states). Moreover, it is well known that the calculations of the response function represents a complex problem even in the ground-state\cite{felix2019,rehr2003,ebert1996,kunes2000,kas2011,liang2017,ikeno2009,shirley2005,woicik2020}; here one needs to go well beyond the ground state and calculate both XAS and MCD response functions transiently (i.e. at several times per material per laser pulse) for an out of equilibrium material in the presence of a pump pulse.

In the present work we overcome this difficulty by extending the implementation of state-of-the-art time dependent density functional theory\cite{RG1984,krieger2015,dewhurst2016} to perform transient L-edge response function calculations. Applying this method to the prototypical cases of Co and Ni we determine the conditions under which the XMCD sum rules can correctly capture magnetic moments in ultrafast spin dynamics.
We find that for experimentally typical pulses these two quite distinct methods of deriving the magnetic moments are in remarkably good agreement, fully validating the experimental tool. However, for very high power density laser light (i.e. short intense pulses, or long pulses of very high fluence) we find that the XMCD sum rules break down, with errors exceeding 50\% in the spin moment. The source of this failure we attribute to the laser light exciting a significant fraction of the charge into highly delocalized states, invalidating the "atomic assumptions" of semi-core to $d$-band transitions upon which the XMCD sum rules are based.


{\it Methodology}: 
The time dependent extension of density functional theory (TD-DFT), is a fully first principles approach that has been shown to accurately describe spin dynamics on femtosecond time scales\cite{dewhurst2018,siegrist2019,steil2020,hofherr2020,clemens2020,chen2019}. In order to calculate the response function at L-edge the deep lying ($\sim$870~eV below the Fermi level) 2$p$ states must be treated on the same footing as the valence states, and thus we employ for our calculations the state-of-the-art all-electron full-potential linearized augmented plane wave method\cite{singh}, as implemented in the Elk code\cite{elk,dewhurst2016}. When treating such low energy localized states (2$p$-states) it is important to include relativistic effects and hence in the present work the mass correction, the Darwin, and the spin-orbit coupling terms are all included in the Hamiltonian. The pump laser enters this Hamiltonian via the time-dependent vector potential, with the initial ground state then time propagated to obtain the time-dependent Pauli spinor Kohn-Sham orbitals (for details see SI and Ref. \onlinecite{dewhurst2016}). In doing so we ensure that 2$p$ states are also a part of the laser induced dynamics, in exactly the same way as the valence states.  These spinor orbitals are then used to calculate the transient densities (magnetization, charge, and current), which gives us the first route to calculate the magnetic moments of a material. The use of DFT ensures that for a given exchange-correlation (XC) potential (we have used adiabatic local density approximation in the present work) these spin moments are exact and can be used as the gold standard for accessing spectrally derived moments.

The transient response function is calculated at each time step by using the ground state orbitals and eigenvalues in conjunction with the effective time-dependent occupation numbers. This method of calculating the transient response function has been studied extensively\cite{clemens2020,dewhurst2020}, and demonstrated to be in very good agreement with experiment.
In addition, we first perform a single shot ground-state $GW$ calculation\cite{gw} to determine the position and width of the deep lying 2$p$ states, which are known to be under-bound by the local/semi-local XC functionals within DFT\cite{sharma2005}. These $GW$ calculations are also used to estimate the broadening of the 2$p$ states.
These corrections are then applied to the transient response functions.

The linear response formalism of the TD-DFT is the used to calculate the transient response function:\cite{RG1984,my-book}
\begin{equation} \label{chi} 
\varepsilon^{-1}(\omega)= 1+\chi_0(\omega)\left[1-(v+f_{\rm xc}(\omega))\chi_0(\omega)\right]^{-1}
\end{equation}
where $\varepsilon$, the fully interacting dielectric tensor, $v$ is the Coulomb potential, $\chi_0$ the non-interacting response function, $f_{\rm xc}$ the exchange-correlation kernel.  Electron-hole correlations, which describe excitonic effects, can be treated by correct choice of this kernel\cite{sharma2011}. The dielectric tensor is calculated by treating all quantities in Eq.~\ref{chi} as complex valued matrices (see SI for details). The off diagonal term of this dielectric tensor then yields the XMCD spectra and the diagonal terms give the XAS\cite{felix2019,dewhurst2020,kunes2000,ebert1996}.

\begin{figure}[t]
\includegraphics[width=\columnwidth, clip]{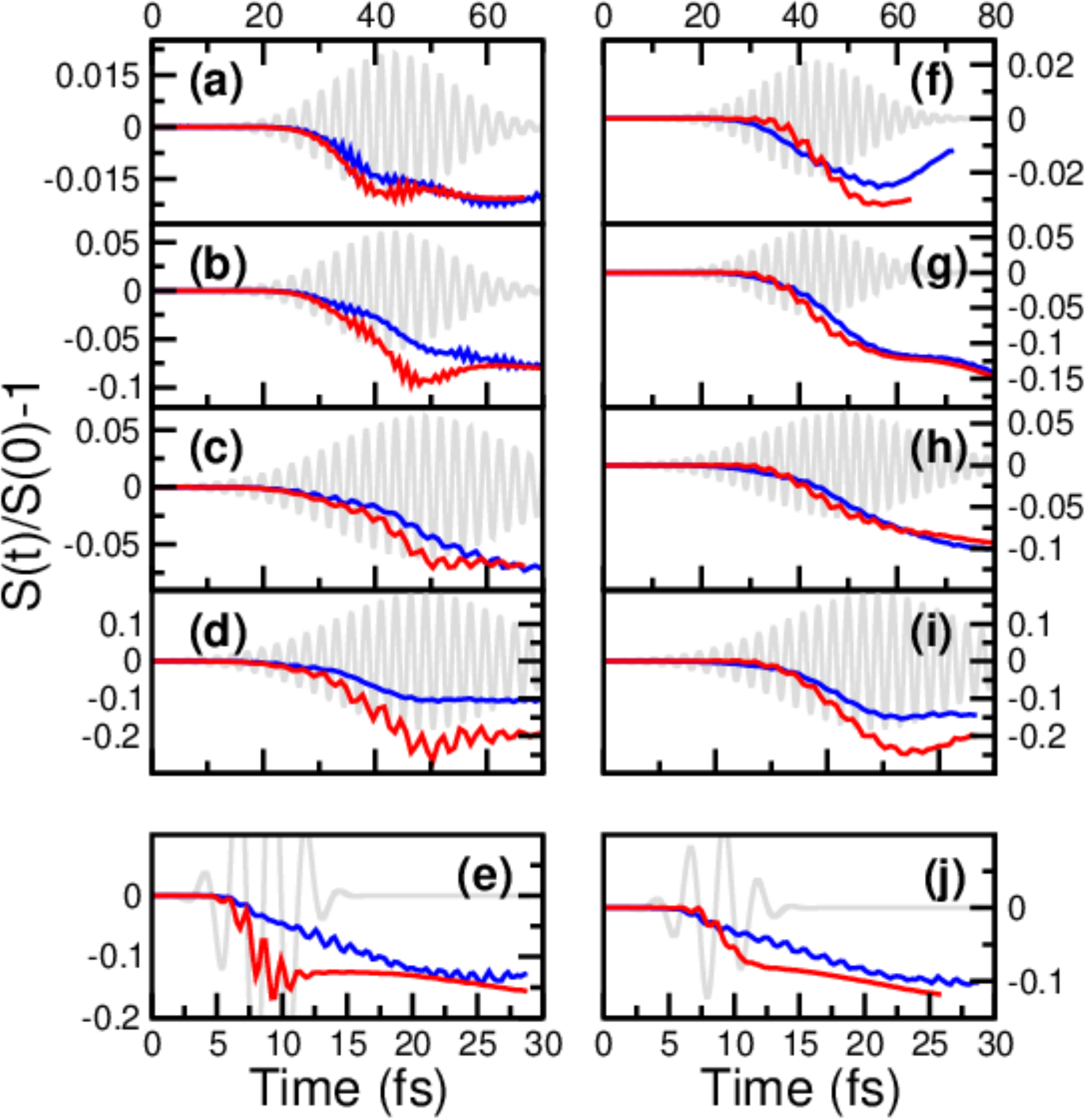}
\caption{Normalized spin angular moment, $S(t)/S(0) -1$ for Ni (a-e) and Co (f-j) pumped with a set of pulses all with central frequency = 1.55eV . The full width half maxima (FWHM), fluence, and pulse intensity in each panel are: (a,f) 24~fs, 2~mJ/cm$^2$, 2x10$^{11}$~W/cm$^2$; (b,g) 24~fs, 6.7~mJ/cm$^2$, 7.2x10$^{12}$W/cm$^2$; (c,h) 36~fs, 6.9~mJ/cm$^2$, 4.8x10$^{11}$~W/cm$^2$; (d,i) 36~fs, 22~mJ/cm$^2$, 2.2x10$^{12}$~W/cm$^2$; (e,j) 5~fs, 5~mJ/cm$^2$, 2.7x10$^{12}$~W/cm$^2$. Results are shown calculated using L-edge sum rules (red) and  using the expectation value of the $\sigma$ operator (blue). The A-field of the pump pulse is also presented (grey). The response function and expectation value derived spin moments can be seen to agree very well, with better agreement for Co than for Ni, and with significant disagreement found only for the most intense laser pulses.
}\label{fig:slp}
\end{figure}

The sum rules\cite{chen1995,thole1992,altarelli1993,resta2020,kunes2000} that provide the relation between the MCD and XAS spectra and the fundamental quantities of spin $\langle S_Z \rangle $ and orbital angular momentum $\langle L_Z \rangle$ are:

\begin{eqnarray}
\left<L_Z\right> &=& (4q)n_h/3r\\ \label{l1}
\left<S_Z\right>&=& (3p-2q)n_h/r
\label{s1}
\end{eqnarray}
where $p$ is integral of the $L_3$ edge (response function obtained due to optical transitions from 2$p_{3/2}$ states to valence band), $q$ is the integral of both the $L_2$ (response function obtained due to optical transitions from 2$p_{1/2}$ states to valence band) and $L_3$ edges, and $r$ is the integral of the XAS over both edges, and $n_h$ is the number of holes in the valence $d$-shell (following common practice we neglect the insignificant spin isotrophy)\cite{chen1995,thole1992,altarelli1993}. In the SI we present the details of $p$, $q$, $r$ and the dependence of the sum rules  upon the energy window in which these are determined.


\begin{figure}[t]
\includegraphics[width=\columnwidth, clip]{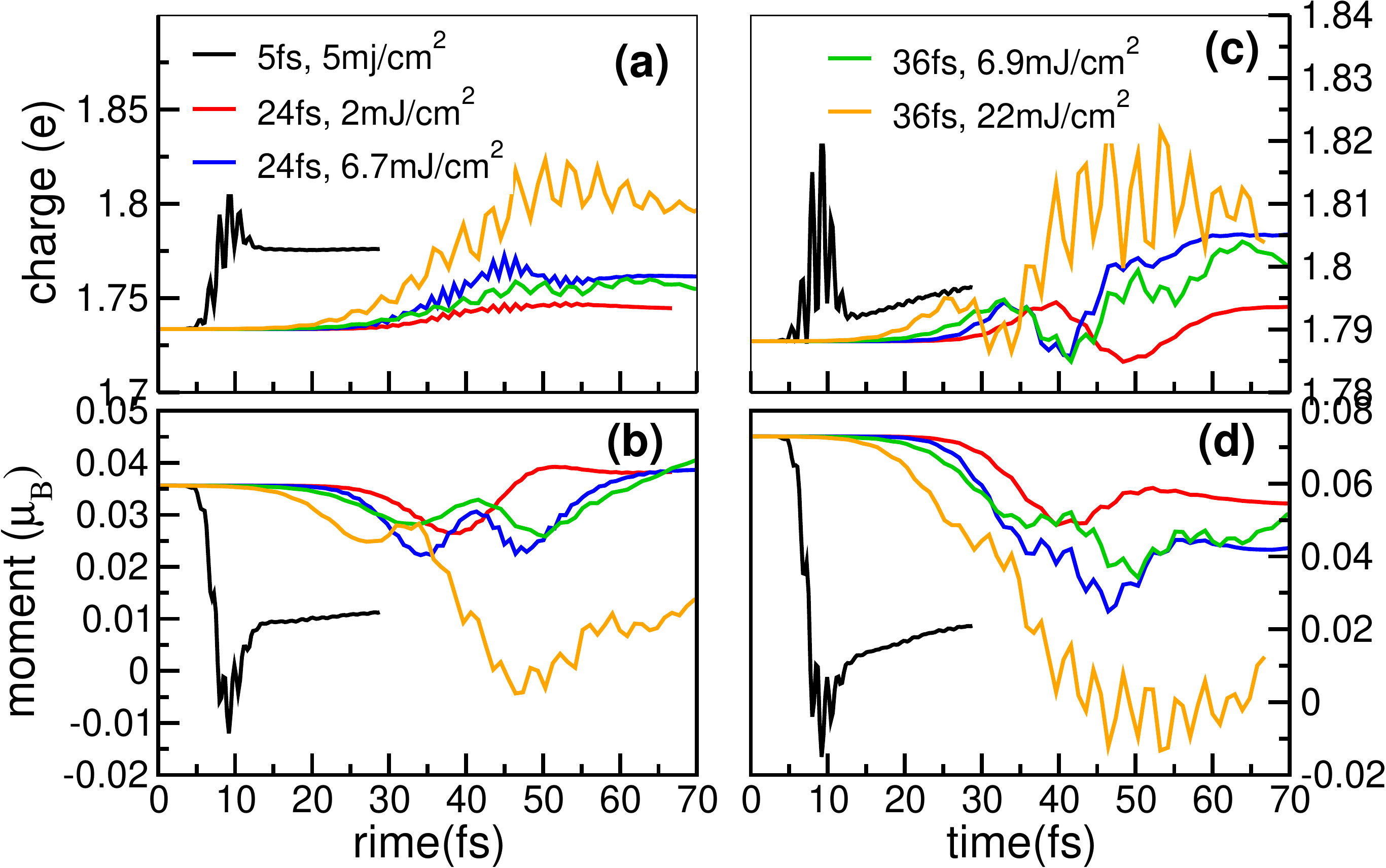}
\caption{Non $d$-band (a) charge and (b) moment for Ni as a function of time (in fs), and (c) charge (d) moment for Co as a function of time. Results are shown for five different laser pulses employed to probe the veracity of the XMCD sum rules, see Fig.~\ref{fig:slp}. A comparison of these figures shows that the non-$d$ moment character of the excited spin correlates very closely with the error of the XMCD sum rule derived moment. Significant errors occur for the most intense pulses, corresponding to a substantial fraction of the excited charge and moment having non-$d$ character.
}
\label{fig:dchg}
\end{figure}

{\it Sum rule for the spin moment}: The L-edge represents the ideal spectral feature for deducing spin information of transition elements: the L$_3$ and L$_2$ absorption edges are well separated in energy (by $\sim$10~eV) allowing unambiguous use of the XMCD sum rules. In Fig.~\ref{fig:slp} we show the spin dynamics in elemental Ni (left column) and Co (right column), with the spin moment calculated both by integrating the magnetization density (the blue lines) as well as derived by from the XMCD spectra using the sum rules, Eq.~\ref{s1} (red lines).
Ideally for a given a XC functional the spin-moment calculated from L-edges should equal to the moment obtained integrating the magnetization density and any deviation can be considered as the inaccuracy of the L-edge for determining the spin-moment. Keeping the full width at half maximum (FWHM) of the pump pulse fixed (24~fs) and increasing the fluence (from 3.55 to 6.7~mJ/cm2) noticeably worsens the accuracy of the spin-moment calculated from sum rules in Ni (compare panels (a) and (b)). 
Fixing the fluence (6.7~mJ/cm$^2$) and increasing the FWHM of the pulse (36~fs) then improved the agreement between the two methods, compare panels (a) and (c). This implies that it is a high power density, i.e. a lot of energy transferred to the system in a short time, that leads to inaccuracy in the moments obtained from XMCD response function at the L-edge.

This can be further tested by reducing the FWHM of the pump pulse to a small value of 5~fs, resulting in a very large amount of energy being pumped into the system in a short time and, as can be seen in Fig.~\ref{fig:slp}(e), this leads to large deviations between the two methods (of the order 68\%). As expected a longer pulse (60~fs) but with a large fluence (22~mJ/cm$^2$) also results in an inaccurate L-edge spin moment.  In the case of Co the accuracy of the L-edge moments is better than for Ni for all the pulses considered-- pulses that result in significant inaccuracy of the L-edge derived moment in Ni (see panels (b) and (g) of Fig.~\ref{fig:slp}) yield good agreement between the two methods for Co. 
We thus conclude that L-edge sum rule derived transient spin moment become unreliable for pump pulses that dump a large amount of energy in the material in a short time. However, this inaccuracy is only during the pulse (most pronounced at the peak of the pulse) and the the two methods come in close agreement after the pulse maximum.
In the present work we have not included the dynamics of the nuclear degrees of freedom and radiative effects thus we follow the demagnetization dynamics in the very early times ($< 60$~fs). We see a demagnetization of the order of 20\% in these very early times.

In the atomic limit, the presence of partly empty $d$-shell would imply that the XUV probe would generate transitions from 2$p$ to 3$d$-states which would, in turn, allow for counting of the empty $d$-states and so the mapping of the spin moment of the material. This is the essence of the XMCD sum rules. However, in solids $d$-states strongly hybridise with $sp$-like states and the pump pulse can thus cause transitions out of the occupied $d$-band and into such $sp$-states. When such an excited system is probed with XUVs, the counting of the empty $d$-states may no longer be indicative of the spin moment of the material. In order to explore the microscopic reason behind the deviation of transient L-edge derived spin moment from the integrated magnetization density, we examine the nature of the excited charge. In Fig.~\ref{fig:dchg} we plot the charge excited to non $d$-states under the influence of various pump laser pulses. 

\begin{figure}[t]
\includegraphics[width=\columnwidth, clip]{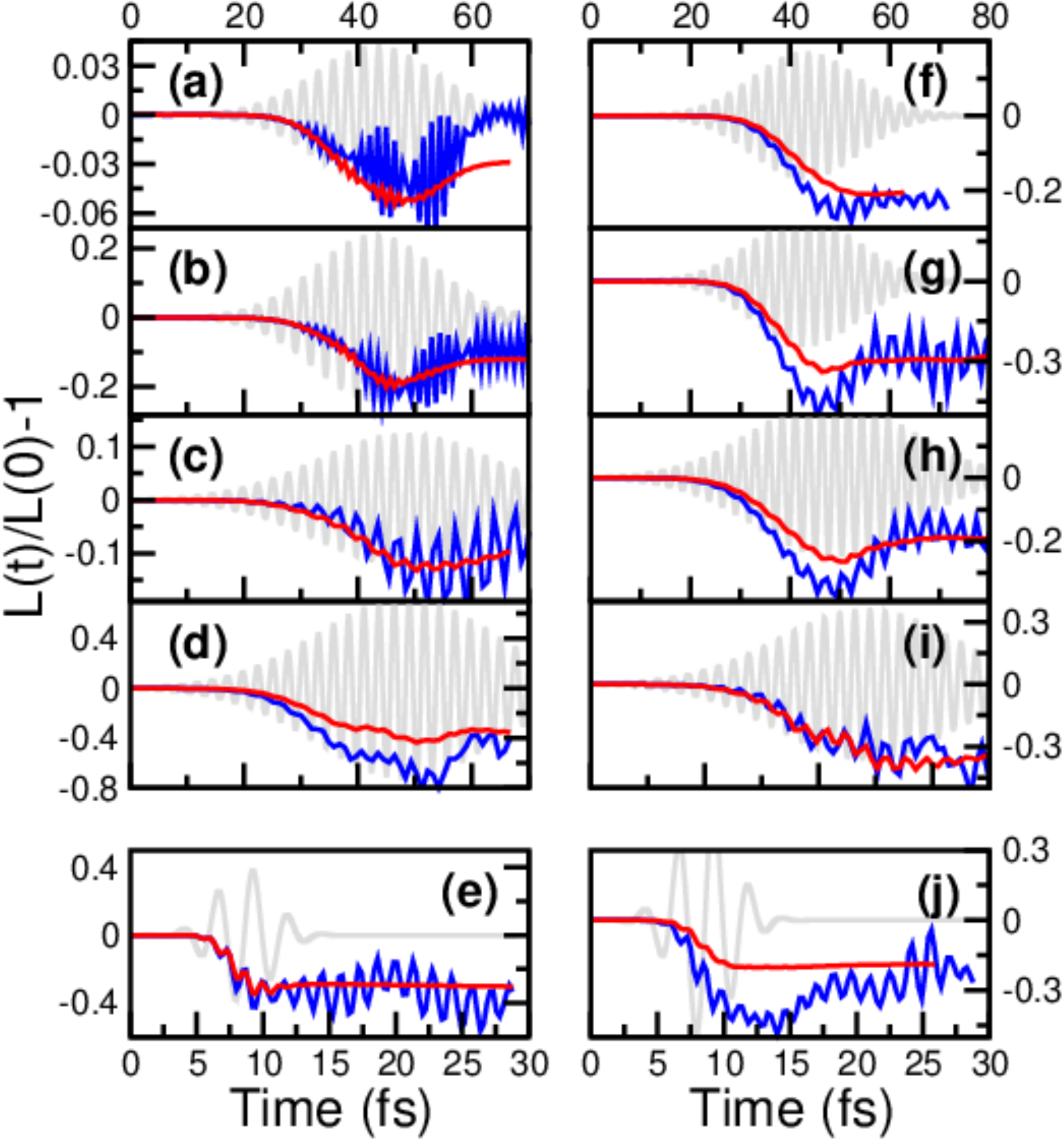}
\caption{Normalized orbital angular moment, $L(t)/L(0) -1$ for Ni (a-e) and Co (f-j) pumped with a set of pulses all with central frequency = 1.55eV. The laser pulses (identical to those employed in Fig.~\ref{fig:slp}) have full width half maxima (FWHM), fluence, and pulse intensity in each panel are: (a,f) 24~fs, 2~mJ/cm$^2$, 2x10$^{11}$~W/cm$^2$; (b,g) 24~fs, 6.7~mJ/cm$^2$, 7.2x10$^{12}$W/cm$^2$; (c,h) 36~fs, 6.9~mJ/cm$^2$, 4.8x10$^{11}$~W/cm$^2$; (d,i) 36~fs, 20~mJ/cm$^2$, 2.2x10$^{12}$~W/cm$^2$; (e,j) 5~fs, 5~mJ/cm$^2$, 2.7x10$^{12}$~W/cm$^2$. Results are calculated using the L-edge sum rule and using the expectation value of the $L=\v r \times \v p$ operator evaluated in the muffin-tin; the A-field of the pump pulse is also presented in grey. Interestingly, while laser induced charge oscillations across the MT boundary render $L$ highly oscillatory, however its time average over a laser cycle agrees closely with the XMCD sum rule derived moment.
}\label{fig:l}
\end{figure}

As can be seen, the breakdown of the XMCD derived spin moment occurs when $d$-band leakage is high. This is especially clear for the two high intensity pulses: (i) the pulse with FWHM 5~fs and fluence 5~mJ/cm$^2$ (see black lines in Fig.~\ref{fig:dchg} and panels (e) and (j) of Fig.~\ref{fig:slp}) displays a clear peak in non-$d$ moment at 10~fs, exactly corresponding to the maximum deviation between XMCD and magnetization density derived moments, and (ii) for the pulse with FWHM 36~fs and fluence 20~mJ/cm$^2$ (orange lines in Fig.~\ref{fig:dchg} and panels (d) and (i) of Fig.~\ref{fig:slp}) the maximum non-$d$ moment is at 50~fs which again corresponds to the maximum deviation between the two methods. For all other pulses the non- $d$ moment is significantly less than these two cases, and as can be seen in Fig. \ref{fig:slp} the error of the L-edge derived moment is correspondingly less.

We thus conclude that the XMCD spin moment sum rule breaks down due to pump laser induced leakage outside the $d$-band. This is predominantly driven by the excitation of charge into high energy delocalized states, as can be seen by the fact that significant error in the spectral moment occurs only during the pump pulse: as excited charge relaxes back to the $d$-band after the pulse envelope has passed the agreement between the two methods improves for all pulses considered (a maximum error exceeding 50\% at pulse peak becomes only 5\% at the end of the simulation window).


{\it Sum rule for the orbital moment}:
The inapplicability of the orbital angular momentum operator $\v L = \v r \times \v p$ to periodic systems is the subject of a large literature\cite{xiao2010}. While a "poormans $L$" can be defined by restricting the evaluation of $\v r \times \v p$ to the spheres centred around each atom, the so called muffin-tins (MT), for laser pumped systems this approach will fail as current loops of excited charge (both open and closed) will inevitably flow between the interstitial and MT regions. The orbital moment calculated from L-edge sum rules, on the other hand, will not suffer from this problem as the Kohn-Sham states used to determine the XMCD spectra are defined over the entire unit cell. As pointed out by Resta\cite{resta2020}, while this approach avoids explicit involvement of the illegitimate $\v r$ operator it too, rigorously speaking, cannot determine the exact orbital-moment in solids. The upshot of all this is that neither approach to calculating the orbital moments is accurate. The source of the errors in these two methods are, however, obviously very different and a good agreement between the two methods would, to some degree, validate the values obtained.

In Fig.~\ref{fig:l} is shown $L$ evaluated in the muffin-tin alongside the corresponding XMCD sum rule derived orbital moment for the same set of pulses used in exploring the performance of the XMCD spin moment sum rule. Interestingly, upon averaging out oscillations due charge flowing across MT boundaries\cite{elliott2016}, the agreement between the L-edge derived orbital moment and the MT value is rather good, suggesting that time averaging the latter provides a reasonable description of the orbital moment in dynamical systems. It is noticeable that while the significant errors in the XMCD sum rule derived spin moment corresponded perfectly with increased non-$d$ character of the excited charge, this correspondence does not hold for the orbital angular momentum.

As in experiments, we see that the dynamics of $L$ and $S$ are very different\cite{boeglin2010,bergeard2014,stamm2010} in that for a fixed pulse change  in normalized orbital angular momentum is always larger than the normalized spin angular momentum, and temporally the change in orbital angular momentum precedes any change in spin-moment (see Figs. \ref{fig:slp} and \ref{fig:l}). This more rapid decrease in $L$ occurs due to optical transition of electrons to excited states, resulting in a change to the charge distribution leading to a change in the orbital angular moment while, as electrons carry their spins with them during direct optical transitions, there is no resulting change in spin moment, $S$. The spin moment  changes only at later times due, for example, to SOC mediated spin flips among other processes.

{\it Conclusions}:
Taking as a prototypical case the ultra-fast spin dynamics in Co and Ni, and by extending TD-DFT to transient L-edge response functions, we have directly calculated spin moments (i) from the sum rules applied to the transient MCD and XAS spectra, and (ii) from the integrated magnetization density. While excellent agreement between these methods is found for the pump pulses typically used in experiments, for high power density pulses this agreement breaks down during the application of the laser pulse, during which there is significant excitation of charge into states of non-$d$ character. This "$d$ band leakage" invalidates the assumption of transitions from semi-core to $d$-orbitals fundamental to the XMCD sum rules and, consequently, the L-edge derived moment is found to significantly overestimate demagnetisation with errors exceeding 50\% seen near the pulse envelope maximum. In the experimental drive towards probing spin dynamics on ever shorter time scales and with intense short pulses, the XMCD sum rules must be applied with caution. However the excellent agreement of XMCD sum rules with underlying magnetic moments, both for what are currently typical experimental pulses, as well as high density non-typical pulses at longer times, provides strong support to this experimental tool and should prompt work exploring the accuracy of the XMCD sum rules in more complex multi-component magnets with exchange at early times between different atomic species as well as between the $L$ and $S$ moments.

{\it Supplementary material}:

\emph{Theoretical Details:}
Time dependent extension of density functional theory (TD-DFT), is a fully first principles approach that has been shown to accurately describe spin dynamics on femtosecond time scales\cite{dewhurst2018,siegrist2019,steil2020,hofherr2020,clemens2020,chen2019}. Underpinning this time dependent extension of DFT is the Runge-Gross theorem\cite{RG1984}, that guarantees for common initial states a one-to-one correspondence between time-dependent external potentials and densities at all later times. This allows the construction of a system of non-interacting particles, chosen to have the same density as that of the interacting system for all times. The many-body wave function of the interacting particles is represented by a Slater determinant of these single-particle orbitals. In the fully non-collinear spin-dependent version of this theory\cite{krieger2015,dewhurst2016} these orbitals are governed by the Pauli equation:

\begin{eqnarray}
i\frac{\partial \psi_j({\bf r},t)}{\partial t}&=&
\Bigg[
\frac{1}{2}\left(-i{\nabla} +\frac{1}{c}{\bf A}_{\rm ext}(t)\right)^2 +v_{s}({\bf r},t)+ \nonumber \\
&&\frac{1}{2c} {\m \sigma}\cdot{\bf B}_{s}({\bf r},t) + \nonumber \\
&&\frac{1}{4c^2} {\m \sigma}\cdot ({\nabla}v_{s}({\bf r},t) \times -i{\nabla})\Bigg]
\psi_j({\bf r},t)
\label{KS}
\end{eqnarray}

where ${\bf A}_{\rm ext}(t)$ is a vector potential representing the applied laser field, ${\m \sigma}$ the vector of Pauli matrices $(\sigma_x,\sigma_y,\sigma_z)$, and $v_{s}({\bf r},t) = v_{\rm ext}({\bf r},t)+v_{\rm H}({\bf r},t)+v_{\rm xc}({\bf r},t)$ the Kohn-Sham (KS) effective potential. This consists of the external potential $v_{\rm ext}$, the classical electrostatic Hartree potential $v_{\rm H}$ and the exchange-correlation (XC) potential $v_{\rm xc}$, for which we have used the adiabatic local density approximation. Similarly the KS magnetic field is given by ${\bf B}_{s}({\bf r},t)={\bf B}_{\rm ext}(t)+{\bf B}_{\rm xc}({\bf r},t)$ in which ${\bf B}_{\rm ext}(t)$ represents the external magnetic field and ${\bf B}_{\rm xc}({\bf r},t)$ the exchange-correlation (XC) magnetic field. The final term of Eq.~\eqref{KS} is the spin-orbit coupling term.

\emph{Response function:} The linear response version of TDDFT reads: 
\begin{equation} \label{chi} 
\chi(\omega)= 1+\chi_0(\omega)\left[1-(v+f_{\rm xc}(\omega))\chi_0(\omega)\right]^{-1}
\end{equation}
where  $v$ is the Coulomb potential, $\chi_0$ the non-interacting response function, $\chi$ is the fully interacting response function, $f_{\rm xc}$ the exchange-correlation kernel. This equation is a matrix equation in reciprocal space vectors {\bf G}, as an external perturbation $e^{i({\bf G}+{\bf q}) \cdot {\bf r}}$ generates a response in the density of the form $e^{i({\bf G'}+{\bf q})\cdot{\bf r}}$. In order to solve this equation for $\chi$ one requires inversion of the matrix $1-(v+f_{\rm xc})\chi_0$ in {\bf G} space. This in turn allows for inclusion of the microscopic components known as the local field effects (LFE). These LFE can be crucial for accurate description of the response function\cite{felix2019}, and in the present work these LFE are included. Furthermore, in order to account of excitonic effects we have used bootstrap approximation\cite{sharma2011} for f$_{\rm xc}$.
From Eq. \ref{chi} one can calculate the dielectric tensor:
\begin{equation}
\varepsilon^{-1}(\omega)= 1+v\chi_(\omega)
\end{equation}
The diagonal of this tensor gives the XAS and the off-diagonal components are used to determine the MCD\cite{felix2019,dewhurst2020,kunes2000}.

\emph{Computational details:} All calculations are performed using the highly accurate full potential linearized augmented-plane-wave method\cite{singh}, as implemented in the ELK\cite{elk} code. 
A smearing width of 0.027~eV was employed for the ground-state as well as for time propagation. For the response function calculations a smearing of 0.9 eV was used. The choice of this latter smearing is based on the ground-state $GW$ calculations: it is the average of the width of p$_{1/2}$ and p$_{3/2}$ states. 
A face centred cubic unit cell with lattice parameter of 3.21\AA for Co 3.53\AA for Ni was used. The Brillouin zone was sampled with a $20\times 20\times 20$ k-point mesh. For time propagation the algorithm detailed in Ref.~\onlinecite{dewhurst2016} was used with a time-step of $2.42$ atto-seconds. 

\emph{Results:} The final magnetization value for each atom is converged with these parameters. We obtained a ground-state spin moment, using the LDA XC functional, of $1.69 \mu_{\rm B}$ for Co and $0.61 \mu_{\rm B}$ for Ni atom. Similar to previous theoretical work we find an orbital moment (calculated from the expectation of $\v r \times \v p$ evaluated in the maximal muffin-tin) of $0.076 \mu_{\rm B}$ for Co and $0.047 \mu_{\rm B}$ for Ni (the corresponding experimental values are $0.14 \mu_{\rm B}$ and $0.053 \mu_{\rm B}$ respectively). As can be seen here, and as is well known, the values of the orbital moment found using the LDA or GGA functionals are smaller than those found in experiment\cite{huhne1998,ebert1996,ceresoli2010,carva2009}. 

\begin{figure}[h]
{\includegraphics[width=0.5\textwidth]{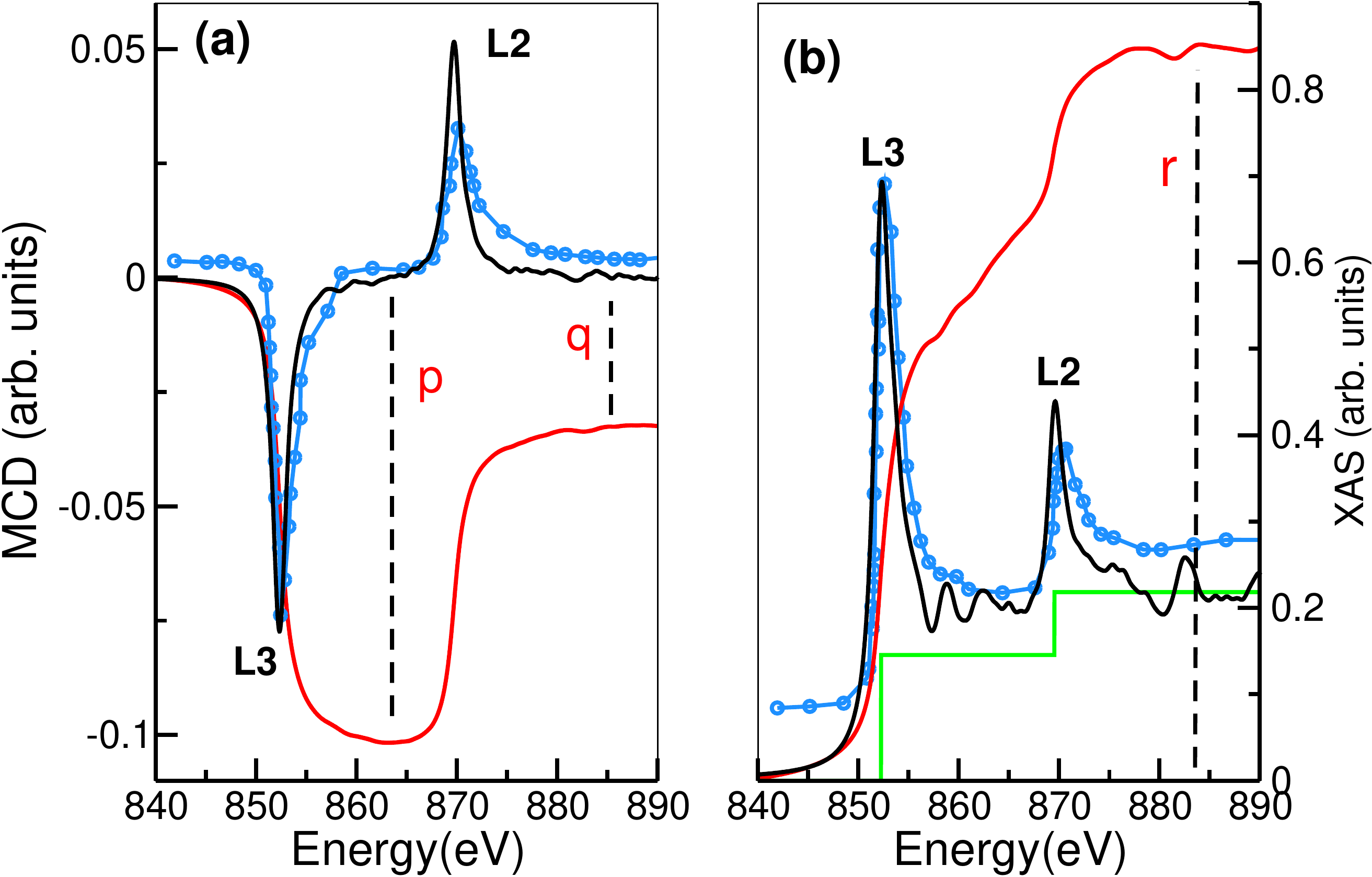}}
    \vspace{0.2cm}
    \caption{(a) MCD and (b) XAS spectra in the ground-state for Ni. Red lines are the integrated quantities (divided by 2 for MCD and 4 by XAS). The green line in (b) is the edge jump removal function subtracted from XAS before integration. Quantities $p$, $q$ and $r$ needed for calculating L$_{\rm z}$ and S$_{\rm z}$ (see Eqs. 2 and 3 of the main text) are marked by vertical dotted lines. Experimental data (blue circles) taken from Ref. \cite{chen1990} is also shown for comparison.}
\label{fig:si-pq}
\end{figure}

The XAS and MCD spectra for Ni are shown in Fig.~\ref{fig:si-pq} and are in overall good agreement with previous experimental data\cite{chen1990}. Similar to previous theoretical simulations\cite{ebert1996,wu1994} we find that the hump like feature in the experimental MCD spectra at around 856~eV is missing from the theoretical result. 
This arises as in the present work we do not account for the core hole effects\cite{woicik2020,shirley2005}, the treatment of which within DFT requires large super-cells, computationally prohibitive for {\it ab-initio} spin dynamics with current computer power. However, it is an interesting question: how do the core-holes impact the transient spectra at the short time scales as studied in the present work, if at all?

The sum rules (Eqs.~2 and 3 of the manuscript) require the quantities $p$, $q$ and $r$ for the calculation of orbital and spin moments, and these quantities are also shown in in Fig.~\ref{fig:si-pq}. These sum rules lead to a spin moment of $0.7 \mu_{\rm B}$ for Ni\cite{huhne1998,wu1994} in the ground-state while the orbital angular momentum turns out to be $0.034 \mu_{\rm B}$.

\begin{figure}[h]
{\includegraphics[width=0.5\textwidth]{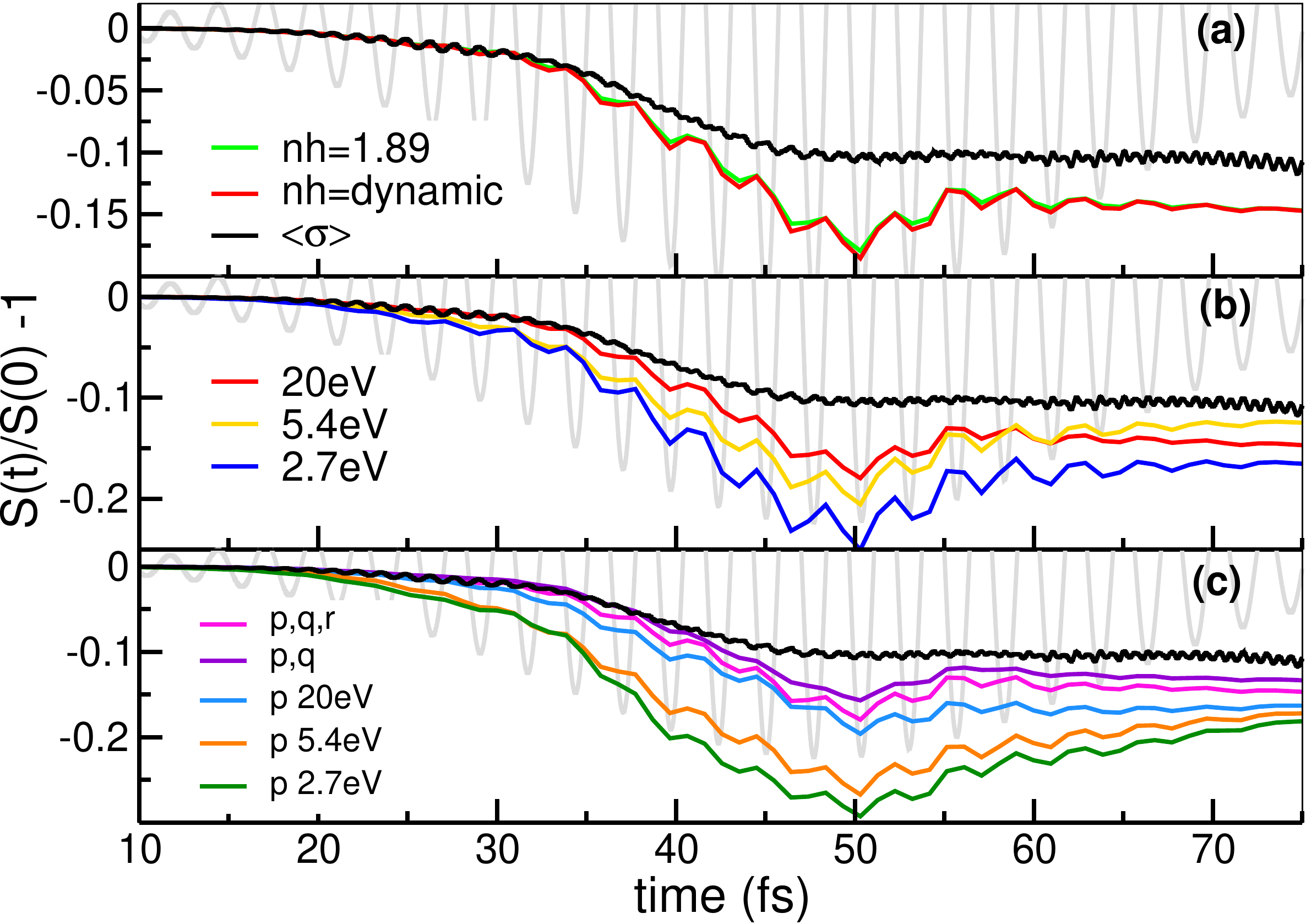}}
    \vspace{0.2cm}
    \caption{Transient normalized moment calculated using various contributions from the XAS and MCD spectra. The ${\v A}$-field of the pump pulse (grey back ground) and the  moment obtained from the magnetization density (black line) are also shown for comparison. Moment calculated using the L-edge sum rules by (a) including dynamical and static ground-state $d$-band holes, (b) calculating $p$, $q$ and $r$ by integrating in various energy windows (20~eV, 5.4~eV and 2.7~eV) around the L3 and L2 edges, and (c) using dynamical $p$, $q$ and $r$; using dynamical $p$ and $q$ but static ground-state value of $r$; using dynamical $p$ obtained by integrating L3 edge in various energy windows (20~eV, 5.4~eV, 2.7~eV) while ignoring the contribution from $q$ and $r$. As can be seen, the most important contribution to the accuracy of L-edge sum rules is the energy window of the probe.}
\label{fig:si-s}
\end{figure}

Experimentally, in order to determine the magnetic moment using the XMCD and XAS spectra often certain approximations are made: (i) the number of holes, $n_h$, (see Eqs.~2 and 3 of the manuscript) is assumed to be static and fixed to the ground-state value, (ii) $p$ and $q$ are calculated from the MCD spectra, but $r$ is assumed to be static, (iii) the normalized moment is assumed to be equal to $p$, which is determined in a small energy window around the L3 edge. In Fig. \ref{fig:si-s} we assess the impact of all these approximations. 

It is clear that the dynamical number of $d$-band holes has no significant effect and one can use the static values (see \ref{fig:si-s}(a)). As expected, increasing the size of the energy window of the probe makes the value of the moment more accurate (\ref{fig:si-s}(b)), taking it closer to the moment obtained using the integrated magnetization density. The impact of using a static value of $r$ instead of determining it from transient XAS is not dramatic (11\% deviation in the worst case). By far the most important source of inaccuracy is the reduced energy window of the probe pulse. From Fig.~\ref{fig:si-s}(c) it is clear that the better the energy resolution the more accurate are the results obtained using the L-edge sum rules.

During the time propagation we see that there are rapid oscillations in the magnetic moment (seen in Figs.~1 and 3 of the manuscript); these are due to the electrons moving back and forth with the frequency of the electric field (as well as higher harmonics). The local moments and charge are extracted by integration of the magnetization and the charge density within a sphere around each atom and this leads to a doubling of the frequency of any oscillation and hence the frequency of these oscillations is twice that of the pump-pulse frequency.

\section{Acknowledgements}
Sharma, SE and CvKS would like to thank DFG for funding through TRR227 (project A04 and A02). Shallcross would like to thank DFG for funding through SH498/4-1 while PE thanks DFG for finding through DFG project 2059421. The authors acknowledge the North-German Supercomputing Alliance (HLRN) for providing HPC resources that have contributed to the research results reported in this paper.

%

\end{document}